\documentclass[nofootinbib,reprint,
longbibliography
]{revtex4-1}
\usepackage{amsmath}
\usepackage{hyperref}
\hypersetup{
	colorlinks = true,
	citecolor = {blue},
	urlcolor = {blue}
}
\usepackage{bm}

\newcommand{\be}{\begin{equation}}
\newcommand{\ee}{\end{equation}}

\newcommand{\zm}{z_\text{m}}
\newcommand{\zh}{z_\text{h}}
\newcommand{\zc}{z_\text{c}}
\newcommand{\xm}{x_\text{m}}
\newcommand{\xc}{x_\text{c}}

\begin{document}


\title{QNEC-Preserving IR Regulators for the Entropy}
\author{Stefan Leichenauer}
\email{sleichen@gmail.com}
\affiliation{Alphabet (Google) X}

\begin{abstract}
Recent work by Ishibashi, Maeda, and Mefford shows that the validity of the QNEC is sensitive to the IR regulator that one uses to define the entropy and its variations. In this note we discuss two general options that preserve both the QNEC and a physically-sensible notion of entropy density. We illustrate the application of each through an example. An important guiding principle is that an IR regulator should have a physical interpretation. 
\end{abstract}


\maketitle

\section{Introduction}\label{sec-intro}

Since it was introduced, the QNEC has led to interesting and surprising insights into both quantum field theory and gravity~\cite{Bousso:2015mna, Bousso:2015wca, Balakrishnan:2017bjg, Leichenauer:2018obf}. The fact that it has been so successful means that we should be wary of situations where it is apparently violated. Recent work by Ishibashi, Maeda, and Mefford~\cite{Ishibashi:2018ern} shows by example that naive attempts to regulate IR divergences in the entropy can lead to tension with the QNEC, and it is the purpose of this note to discuss how this tension can be relieved. As in~\cite{Ishibashi:2018ern}, we will work in the holographic context where the issues are relatively simple to understand, but analogous considerations should hold more broadly.

The holographic proofs of the QNEC generally rely on two facts~\cite{Koeller:2015qmn, Akers:2016ugt, Fu:2017evt, Akers:2017ttv}. The first is Entanglement Wedge Nesting (EWN), which is a manifestation of boundary causality applied to entanglement wedges and is not subject to IR divergences. Indeed, the calculations of~\cite{Ishibashi:2018ern} showed no violation of EWN.

The second fact is the identification of the first variation of the entropy with a term in the small-$z$ expansion of the extremal surface embedding function, where $z$ is the standard Fefferman-Graham bulk coordinate. This identification requires a careful understanding of the divergences associated with the entropy (both UV and IR), and its precise form depends on the spacetime dimension, the bulk gravitational theory, and whether the boundary is curved or flat. For an odd number $d$ of boundary dimensions and Einstein gravity in the bulk, once the appropriate conditions for UV-finiteness have been enforced, the answer is simply~\cite{Akers:2017ttv}
\be\label{eq-Avar}
\frac{1}{\sqrt{h}} \frac{\delta \mathcal{A}}{\delta X^\mu} = - d g_{\mu\nu} X_{(d)}^\nu,
\ee
where $\mathcal{A}$ is the area of the extremal surface in the bulk, $h$ is the intrinsic metric of the entangling surface, $g$ is the metric of the boundary spacetime, $X_{(d)}^\mu$ is the coefficient of $z^d$ in the small-$z$ expansion of the extremal surface coordinate, and the variation is taken with respect to the coordinate position of the entangling surface at a point. It is~\eqref{eq-Avar} that was apparently violated in~\cite{Ishibashi:2018ern} because of issues with IR divergences.

The goal of this note is to show that there are natural IR regulators which preserve \eqref{eq-Avar} and hence also the QNEC. A crucial principle is that a regulator should be a physical way of dealing with divergences, not just a mathematical trick. We will state the options now and define them more carefully below:\footnote{The authors of~\cite{Ishibashi:2018ern} propose to cure IR divergences by subtracting infinite constants from the area which ``share the same divergence." Such a definition implicitly involves the use of some regulator which allows one to assess whether the divergences are shared.}
\begin{itemize}
\item {\bf Option 1:} Make the region finite.
\item {\bf Option 2:} Put the system in a box.
\end{itemize}

Before continuing on, it is also worth noting that \eqref{eq-Avar} is of general interest besides its utility in proving the QNEC. It essentially identifies $X_{(d)}^\mu$ as an entropy current. In Section~\ref{sec-example} we will see that this current successfully captures the thermal entropy density in the appropriate limit, which is a nice consistency check.


\section{Generalities}\label{sec-gen}

In this section we'll discuss the derivation of~\eqref{eq-Avar}, the potential issues associated with IR divergences, and how those issues can be dealt with. In the general setup, we choose an entangling surface $\Sigma$ on the boundary and are instructed to find the bulk surface extremizing some appropriate entropy functional~\cite{Ryu:2006bv, Hubeny:2007xt, Dong:2013qoa, Engelhardt:2014gca}.

Consider the case where the entropy functional is of the form\footnote{We could be even more general, but the notation would get clumsy.}
\be
S = \int_\Omega \mathcal{L}(X,\partial X),
\ee
where $X^\mu(\sigma)$ are the embedding functions for the bulk surface and $\Omega$ represents the integration domain of the internal coordinates $\sigma$. The entropy contains divergences---both UV and IR---that we can regulate by restricting the domain of integration to $\Omega' \subset \Omega$, later taking the limit $\Omega'\to \Omega$. Divergences in the entropy are associated to noncompactness of the extremal surface, so in defining $\Omega'$ we should excise only those portions of $\Omega$ which map to infinity in the bulk. Extra unnecessary excisions complicate the discussion and will be assumed absent.

Shape deformations of $\Sigma$ involve modifying the functions $X^\mu(\sigma)$ so that they have new boundary conditions but still satisfy the equations of motion. Then the calculus of variations tells us that\footnote{In this section we are choosing internal coordinates so that $\Omega'$ is invariant under shape deformations.}
\be
\delta S = \lim_{\Omega'\to \Omega} \int_{\partial\Omega'} \frac{\partial \mathcal{L}}{\partial (\partial_i X^\mu)}n_i\delta X^\mu,
\ee
where $n_i$ is normal to $\partial\Omega'$.

In the limit $\Omega \to \Omega'$, the image of $\partial \Omega'$ will, by construction, limit to a union of points at infinity from the bulk point of view. Some of these points will constitute the original entangling surface $\Sigma$, while others may reach spatial infinity without returning to the boundary. This can happen, for example, if the extremal surface asymptotes to a noncompact black hole horizon. Those points are associated with IR divergences of the entropy.

By a slight abuse of notation, we can write the variation of the entropy as
\be\label{eq-twoterms}
\delta S =  \int_{\Sigma} \frac{\partial \mathcal{L}}{\partial (\partial_i X^\mu)}n_i\delta X^\mu + \int_{\Sigma'} \frac{\partial \mathcal{L}}{\partial (\partial_i X^\mu)}n_i\delta X^\mu,
\ee
where $\Sigma'$ represents those parts of $\partial\Omega'$ which do not limit to $\Sigma$. The integration over $\Sigma$ reproduces~\eqref{eq-Avar}---once the UV divergences are accounted for~\cite{Akers:2017ttv}---while the integration over $\Sigma'$ is unwanted.

Defining the integration over $\Sigma'$ carefully, or otherwise dealing with it, is the job of an IR regulator. There are two obvious options, which we mentioned above and will now define:

\begin{itemize}
\item {\bf Make the region finite.} Simply put, replace the entangling surface $\Sigma$ by another one, $\bar{\Sigma}$, which bounds a compact region and by construction does not have any IR divergences in the entropy. Do all calculations for $\bar{\Sigma}$, then take limit where $\bar{\Sigma} \to \Sigma$. Since~\eqref{eq-Avar} holds by construction for $\bar{\Sigma}$ it will remain true in the limit.
\item {\bf Put the system in a box.} Another possibility, which serves to regulate all IR divergences, including those of the entropy, is to place the entire system inside a large box and then take the size of the box to infinity at the end. More precisely, this means placing the field theory on a manifold with boundary (the ``box"). The gravity dual of such a setup was discussed in~\cite{Takayanagi:2011zk,Fujita:2011fp}, where it was argued that the walls of the box extend into the bulk in the form of a brane with certain boundary conditions. When calculating the entropy, one allows the extremal surface terminate on the brane in whatever way minimizes the entropy functional. We will not attempt a general analysis of this setup here, but we will see below in an example that the net effect is that the second term in~\eqref{eq-twoterms} is eliminated as the IR regulator is taken to infinity.
\end{itemize}

In the remainder of this note we will work through an explicit example of a system with IR divergences in the entropy, and show how to regulate it using both of these methods.


\section{An Example}\label{sec-example}

An illustrative example is a cylindrical black hole in the bulk with metric
\begin{align}
ds^2 = \frac{1}{z^2}\biggl[&-(1-(z/\zh)^3)dt^2\nonumber\\ 
&+ \frac{dz^2}{1-(z/\zh)^3} + dx^2 + d\phi^2\biggr].
\end{align}
We choose our boundary region to be the strip $x_0 < x < x_1$ at some fixed time, so that the entangling surfaces are circles wrapping the $2\pi$-periodic $\phi$ direction at $x = x_0$ and $x=x_1$. The region with IR divergences that we are trying to regulate is the half-space $x>x_0$, obtained by setting $x_1 = \infty$. This geometry was considered in~\cite{Ishibashi:2018ern}, and is qualitatively similar to the numerical example they constructed to violate the QNEC.
 

\subsection{Make the Region Finite}

First, we will consider the half-space in terms of the limit $x_1\to \infty$. For finite $x_1$, we can write the extremal area functional as
\be
\mathcal{A} = 4\pi \int_{0}^{\zm} \frac{dz}{z^2} \sqrt{x'^2+\frac{1}{1- (z/\zh)^3}}~,
\ee
where $\zm < \zh$ is the maximal value of $z$ that the surface reaches. The function $x(z)$ satisfies
\be
x'(z)= \frac{(z/\zm)^2}{\sqrt{(1-(z/\zm)^4)(1- (z/\zh)^3)}}~,
\ee
from which we find the near-boundary expansion
\be\label{eq-exp}
x(z) = x_0 + \frac{1}{3}\frac{z^3}{\zm^2} + \cdots.
\ee

Variations of the area with respect to $x_0$, for fixed $x_1$, are easily obtained using the calculus of variations.\footnote{When we vary $x_0$ here we are varying $x_0$ at all values of $\phi$ simultaneously. In the business of shape variations it is very important to distinguish this kind of non-local variation from more general local variations. For our present purposes these simple nonlocal variations are enough.} We find 
\be\label{eq-varAfin}
\delta \mathcal{A} = -\frac{4\pi}{\zm^2} \left( \delta x_0 - \delta \xm \right)  = -\frac{2\pi}{\zm^2} \delta x_0.
\ee
Here $\delta \xm$ refers to the change in the $x$ coordinate of the surface at $z=\zm$ (keeping in mind that $\zm$ also varies with the surface). When $x_1$ is fixed, $\delta\xm = \delta x_0/2$ by symmetry and so we get our final answer. Happily, using~\eqref{eq-exp} we see that this matches the integral of~\eqref{eq-Avar} over the $\phi$ circle.

We'll also note that this answer has a very sensible physical interpretation in terms of entropy density. In the limit $x_1\to\infty$, we find that $\delta S = -2\pi \delta x_0/4G\zh^2 = -2\pi s \delta x_0$, where $s$ is the thermal entropy density of the state. So we would conclude that the entropy of the half-space $x>x_0$ changes by an amount equal to the gained or lost thermal entropy as a result of moving $x_0$.


\subsection{Put the System in a Box}

Now let's consider setting $x_1 = \infty$ from the beginning. Then the area functional is 
\be\label{eq-infA1}
\mathcal{A} = 2\pi \int_{0}^{\zh} \frac{dz}{z^2} \sqrt{x'^2+\frac{1}{1- (z/\zh)^3}}~,
\ee
and the solutions satisfy 
\be\label{eq-xprime}
x'(z) = \frac{(z/\zh)^2}{\sqrt{(1-(z/\zh)^4)(1- (z/\zh)^3)}}~.
\ee
Formally speaking, we can plug this back in to the area functional to find
\be\label{eq-infA}
\mathcal{A} = 2\pi \int_{0}^{\zh} \frac{dz}{z^2} \frac{1}{\sqrt{(1-(z/\zh)^4)(1- (z/\zh)^3)}}~,
\ee
which now seems to have no dependence on $x_0$ at all, and hence would have zero variation. But since the area is IR divergent, we really should regulate it first before asking about its variation.

The regulator we consider now is to place the field theory on a manifold with boundary (the ``box") and take the limit of a very large box. The wall of the box is located at $x= L$, and we will take $L \to \infty$ at the end. In~\cite{Takayanagi:2011zk,Fujita:2011fp} it was argued that in the bulk there should be a brane anchored at $x=L$, $z=0$, that extends into the bulk and effectively cuts off the bulk geometry. Extremal surfaces are allowed to end on the brane in whatever way minimizes the area. The brane may bend, which means that the position of the bulk cutoff will follow a trajectory $x = \xc(z)$. We will not assume much about this trajectory, other than that $\xc \sim L$ for all $z<\zh$.
  
The net effect of this regulator is that the integration range of~\eqref{eq-infA1} is terminated at $\zc < \zh$, where $\zc$ is the value of $z$ where the extremal surface reaches $x=\xc$. Using the calculus of variations one last time, we find
\be\label{eq-varAbox}
\delta \mathcal{A} = -\frac{2\pi}{\zh^2} \left( \delta x_0 - \delta \xc \right) + 2\pi\frac{1}{\zc^2}\sqrt{\frac{1-(\zc/\zh)^4}{1- (\zc/\zh)^3}}\delta \zc~.
\ee
In the language of Section~\ref{sec-gen}, the $\delta x_0$ term is the $\Sigma$ term of~\eqref{eq-twoterms} and the $\delta\xc$ and $\delta\zc$ terms come from $\Sigma'$. We will see that the latter drop out as $L\to \infty$.

The ratio $\delta \xc/\delta \zc$ is determined by the shape of the brane at the point of intersection with the extremal surface. As $L\to \infty$, we have $\zc \to \zh$ and this ratio should remain finite in that limit (it will be equal to zero if the brane sits at a constant $\xc = L$). Another constraint comes from the shape of the extremal surface, namely that the difference $\xc - x_0$ should be obtained by integrating~\eqref{eq-xprime}. This implies that
\be
\delta \xc - \delta x_0 = x'(\zc) \delta \zc.
\ee 
As $\zc \to \zh$ we have $x'(\zc) \to \infty$, which means $\delta \xc$, and consequently $\delta \zc$, both go to zero. As a result, both of the unwanted terms in~\eqref{eq-varAbox} disappear and we reproduce~\eqref{eq-varAfin}.

Equation~\eqref{eq-varAbox} is sufficiently general that we can use it to analyze the effects of an unmotivated, unphysical IR regulator as well, just to see what goes wrong. Suppose we place an imaginary cutoff surface at fixed $z=\zc$, independent of $x$. Then $\xc$ in \eqref{eq-varAbox} represents the value of $x$ at which the extremal surface intersects $\zc$. Then $\delta \zc = 0$ by definition and $\delta \xc = \delta x_0$ by symmetry, so we find the naive answer $\delta \mathcal{A} =0$. However, the prescription to fix $\zc$ is not well-motivated, and also not a very general way to deal with divergences. If we decompactify the $\phi$ direction, for instance, then another IR divergence will appear that cannot be tamed this way.

\section{Discussion}

We have argued for two physically meaningful IR regulators for the entropy that preserve~\eqref{eq-Avar}, which in turn allows one to prove the QNEC and gives rise to a natural physical interpretation of the thermal entropy density. The first regulator involved making the entangling region finite before calculating shape variations of the entropy, and then letting the size of the region go to infinity at the end. The second was to place the system in a box, which is a familiar way to regularize IR divergences in field theory. That prescription has the advantage of regulating all IR divergences, not just for the entropy, but has the disadvantage of being more technically involved due to the issue of boundary conditions on the box and, in the holographic case, the dynamics of the bounding brane. We did not attempt to provide a general proof that placing the system in a box preserved~\eqref{eq-Avar}, but we showed how it worked in a particular example and the general case is likely similiar.


\begin{acknowledgments}
I would like to thank V.~Chandrasekaran, A.~Levine, and A.~Shahbazi-Moghaddam for bringing~\cite{Ishibashi:2018ern} to my attention and discussing it. Thank you as well to C.~Akers and R.~Bousso for helpful comments on an earlier draft.
\end{acknowledgments}


\bibliographystyle{utcaps}
\bibliography{QNEC-IR-regulator}

\providecommand{\href}[2]{#2}\begingroup\raggedright\begin{thebibliography}{10}

\bibitem{Bousso:2015mna}
R.~Bousso, Z.~Fisher, S.~Leichenauer, and A.~C. Wall, ``{Quantum focusing
  conjecture},'' \href{http://dx.doi.org/10.1103/PhysRevD.93.064044}{{\em Phys.
  Rev.} {\bf D93} (2016) no.~6, 064044},
\href{http://arxiv.org/abs/1506.02669}{{\tt arXiv:1506.02669 [hep-th]}}.

\bibitem{Bousso:2015wca}
R.~Bousso, Z.~Fisher, J.~Koeller, S.~Leichenauer, and A.~C. Wall, ``{Proof of
  the Quantum Null Energy Condition},''
\href{http://arxiv.org/abs/1509.02542}{{\tt arXiv:1509.02542 [hep-th]}}.

\bibitem{Balakrishnan:2017bjg}
S.~Balakrishnan, T.~Faulkner, Z.~U. Khandker, and H.~Wang, ``{A General Proof
  of the Quantum Null Energy Condition},''
\href{http://arxiv.org/abs/1706.09432}{{\tt arXiv:1706.09432 [hep-th]}}.

\bibitem{Leichenauer:2018obf}
S.~Leichenauer, A.~Levine, and A.~Shahbazi-Moghaddam, ``{Energy is
  Entanglement},''
\href{http://arxiv.org/abs/1802.02584}{{\tt arXiv:1802.02584 [hep-th]}}.

\bibitem{Ishibashi:2018ern}
A.~Ishibashi, K.~Maeda, and E.~Mefford, ``{Violation of the QNEC in a
  holographic wormhole and IR effects},''
\href{http://arxiv.org/abs/1808.05192}{{\tt arXiv:1808.05192 [hep-th]}}.

\bibitem{Koeller:2015qmn}
J.~Koeller and S.~Leichenauer, ``{Holographic Proof of the Quantum Null Energy
  Condition},'' \href{http://dx.doi.org/10.1103/PhysRevD.94.024026}{{\em Phys.
  Rev.} {\bf D94} (2016) no.~2, 024026},
\href{http://arxiv.org/abs/1512.06109}{{\tt arXiv:1512.06109 [hep-th]}}.

\bibitem{Akers:2016ugt}
C.~Akers, J.~Koeller, S.~Leichenauer, and A.~Levine, ``{Geometric Constraints
  from Subregion Duality Beyond the Classical Regime},''
\href{http://arxiv.org/abs/1610.08968}{{\tt arXiv:1610.08968 [hep-th]}}.

\bibitem{Fu:2017evt}
Z.~Fu, J.~Koeller, and D.~Marolf, ``{The Quantum Null Energy Condition in
  Curved Space},'' \href{http://dx.doi.org/10.1088/1361-6382/aa8f2c,
  10.1088/1361-6382/aaa5c9}{{\em Class. Quant. Grav.} {\bf 34} (2017) no.~22,
  225012}, \href{http://arxiv.org/abs/1706.01572}{{\tt arXiv:1706.01572
  [hep-th]}}.
[Erratum: Class. Quant. Grav.35,no.4,049501(2018)].

\bibitem{Akers:2017ttv}
C.~Akers, V.~Chandrasekaran, S.~Leichenauer, A.~Levine, and
  A.~Shahbazi-Moghaddam, ``{The Quantum Null Energy Condition, Entanglement
  Wedge Nesting, and Quantum Focusing},''
\href{http://arxiv.org/abs/1706.04183}{{\tt arXiv:1706.04183 [hep-th]}}.

\bibitem{Ryu:2006bv}
S.~Ryu and T.~Takayanagi, ``{Holographic derivation of entanglement entropy
  from AdS/CFT},'' \href{http://dx.doi.org/10.1103/PhysRevLett.96.181602}{{\em
  Phys. Rev. Lett.} {\bf 96} (2006)  181602},
\href{http://arxiv.org/abs/hep-th/0603001}{{\tt arXiv:hep-th/0603001
  [hep-th]}}.

\bibitem{Hubeny:2007xt}
V.~E. Hubeny, M.~Rangamani, and T.~Takayanagi, ``{A Covariant holographic
  entanglement entropy proposal},''
  \href{http://dx.doi.org/10.1088/1126-6708/2007/07/062}{{\em JHEP} {\bf 07}
  (2007)  062},
\href{http://arxiv.org/abs/0705.0016}{{\tt arXiv:0705.0016 [hep-th]}}.

\bibitem{Dong:2013qoa}
X.~Dong, ``{Holographic Entanglement Entropy for General Higher Derivative
  Gravity},'' \href{http://dx.doi.org/10.1007/JHEP01(2014)044}{{\em JHEP} {\bf
  01} (2014)  044},
\href{http://arxiv.org/abs/1310.5713}{{\tt arXiv:1310.5713 [hep-th]}}.

\bibitem{Engelhardt:2014gca}
N.~Engelhardt and A.~C. Wall, ``{Quantum Extremal Surfaces: Holographic
  Entanglement Entropy beyond the Classical Regime},''
  \href{http://dx.doi.org/10.1007/JHEP01(2015)073}{{\em JHEP} {\bf 1501} (2015)
   073},
\href{http://arxiv.org/abs/1408.3203}{{\tt arXiv:1408.3203 [hep-th]}}.

\bibitem{Takayanagi:2011zk}
T.~Takayanagi, ``{Holographic Dual of BCFT},''
  \href{http://dx.doi.org/10.1103/PhysRevLett.107.101602}{{\em Phys. Rev.
  Lett.} {\bf 107} (2011)  101602},
\href{http://arxiv.org/abs/1105.5165}{{\tt arXiv:1105.5165 [hep-th]}}.

\bibitem{Fujita:2011fp}
M.~Fujita, T.~Takayanagi, and E.~Tonni, ``{Aspects of AdS/BCFT},''
  \href{http://dx.doi.org/10.1007/JHEP11(2011)043}{{\em JHEP} {\bf 11} (2011)
  043},
\href{http://arxiv.org/abs/1108.5152}{{\tt arXiv:1108.5152 [hep-th]}}.

\end{thebibliography}\endgroup

\end{document}